\documentclass[amssymb,aps,prl,twocolumn,groupedaddress,showpacs,floatfix]{revtex4}
\usepackage{amsmath}
\usepackage{epsfig}

\begin{document}

\title{Purely orbital diamagnetic to paramagnetic fluctuation \\
of quasi two-dimensional carriers
under in-plane magnetic field}

\author{Constantinos Simserides
\footnote{http://www.matersci.upatras.gr/simserides}}

\affiliation{
University of Patras, Materials Science Department, Panepistimiopolis, Rio,
GR-26504, Patras, Greece and
Institute of Materials Science, National Center of Scientific Research Demokritos,
GR-15310 Athens, Greece.}

\date{\today}

\begin{abstract}
An external magnetic field, $H$, applied parallel to
a quasi two-dimensional system modifies
quantitatively and qualitatively the density of states.
Using a self-consistent numerical approach,
we study how this affects the entropy, $S$,
the free energy, $F$, and the magnetization, $M$,
for different sheet carrier concentrations, $N_s$.
As a prototype system we employ III-V double quantum wells.
We find that although $M$ is mainly in the opposite direction of $H$,
the system is not linear.
Surprisingly $\partial M / \partial H$
swings between negative and positive values,
i.e., we predict an entirely orbital diamagnetic to paramagnetic fluctuation.
This phenomenon is important
compared to the ideal de Haas-van Alphen effect
i.e. the corresponding phenomenon
under perpendicular magnetic field.
\end{abstract}



\maketitle

The scrutiny of quasi two-dimensional (2D) carriers
under magnetic field has a long, fruitful history.
Emphasis was given to the configuration
where the magnetic field, $H$, is applied perpendicularly
to the quasi 2D system,
leading to quantization of the free in-plane motion into Landau levels \cite{landau}.
In this configuration
the integer quantum Hall effect \cite{IQHE} was discovered
and the fractional quantum Hall effect was observed \cite{FQHE}
and explained in terms of
quasiparticles with fractional charge \cite{laughlin}.
Besides, varying $H$,
oscillations of the magnetic susceptibility
(the de Haas - van Alphen effect \cite{dehaas-vanalphen})
as well as
oscillations of the conductivity
(the Shubnikov-de Haas effect \cite{shubnikov-dehaas})
were observed.
The heart of the system, the density of states (DOS)
can be directly probed by
measurements of a thermodynamic quantity
like the magnetization, $M$.
Experimental evidence of the ideal de Haas-van Alphen effect
in a quasi 2D carrier system was only found recently
by Wilde et al. \cite{wilde_et_al_2006}
who measured the magnetization oscillations
of {\it high mobility} 2D electrons
in modulation-doped AlGaAs/GaAs heterostructures.
In the sample with the highest oscillation amplitude
they observed discontinuous jumps in $M$
with peak-to-peak amplitude
of two effective Bohr magnetons
($\mu_B^{*} = \frac{e \hbar}{2 m^*}$, $m^*$ is the effective mass)
per electron,
in agreement with old Peierls' prediction \cite{peierls}.
Numerical simulations \cite{wilde_et_al_2006}
assuming no states between the Landau levels (no ``background'' DOS)
could model these jumps quantitatively,
but measurements on samples with lower mobility
revealed a finite background DOS.
Indeed in earlier studies \cite{earlier} a considerable DOS
between the broadened Landau levels was necessary
to simulate the experimental results.

This article is devoted to
the entropy, $S$, the free energy, $F$, the magnetization, $M$,
and the magnetic susceptibility, $\chi_m = \partial M / \partial H$,
of quasi 2D carriers
under {\it in-plane} magnetic field,
employing a self-consistent envelope function approach.
In particular, we show that although $M$
remains basically in the opposite direction of $H$,
the system is highly non-linear.
As a result, the magnetic susceptibility
oscillates between negative and positive values.
This is the first prediction of a purely orbital
diamagnetic to paramagnetic fluctuation.
The effect, ignored by the community up to now,
is important compared to the ideal de Haas-van Alphen effect,
the corresponding phenomenon
under perpendicular magnetic field.
We hope that this article will also help the interpretation
of magnetization measurements under tilted $H$,
i.e., whenever an in-plane component of $H$ exists.

When a quasi 2D system is subjected to
an in-plane -or even tilted- magnetic field,
the charming concept of Landau levels must be revised,
because
carriers move under the competing influence
of the Lorentz force and the force due to the quantum well (QW) confining potential.
The equal-energy surfaces \cite{sj}
or equivalently the density of states \cite{lyo:94,simserides:99}
are qualitatively and quantitatively modified
because the spatial and the magnetic confinement compete.
Generally, a proper treatment involves self-consistent computation
\cite{sj,simserides:99,simserides-prb-2004}
of the energy dispersion, $E_{i,\sigma}(k_x)$, where
$i$ is the subband index,
$\sigma$ denotes the spin
and $k_x$ is the in-plane wave vector
perpendicular to the external in-plane magnetic field (applied along $y$), $H$.
The envelope functions along the ``growth'' $z$-axis depend on $k_x$ i.e.,
$\psi_{i,\sigma,k_x,k_y}({\bf r})
\propto \zeta_{i,\sigma,k_x}(z) e^{i k_x x} e^{i k_y y}$.
The consequences of this modification were initially realized in
transport \cite{transport} experiments.
This amendment also influences
the character of plasmons
in single \cite{plasmons-single} and double \cite{plasmons-double} QWs.
The $\cal N$-type kink was theoretically predicted \cite{PL-theo}
and recently verified in photoluminescence experiments \cite{PL-expe}.
The impact of the DOS modification on various properties of
dilute-magnetic-semiconductor (DMS) single QWs
was studied lately \cite{simserides-prb-2004,simserides-prb-2007-etc}.
Thus, it seems that the parallel or titled configuration
offers new avenues to explore.
A compact formula for
the DOS of a quasi 2D system under in-plane $H$
exists \cite{simserides-prb-2004}:

\begin{equation}\label{dos}
\rho({\mathcal E}) = \frac {A \sqrt{2m^*}}{4 \pi^2 \hbar}
\sum_{i,\sigma} \int_{-\infty}^{+\infty} \! dk_x
\frac{\Theta({\mathcal E}-E_{i,\sigma}(k_x))}
{ \sqrt {{\mathcal E}-E_{i,\sigma}(k_x)} }.
\end{equation}

\noindent It is implied that the QWs are along the $z$-axis
and $H$ is applied along the $y$-axis.
$\Theta$ is the step function, $A$ is the $xy$-area
of the structure.
$E_{i,\sigma}(k_x)$ are the spin-dependent $xz$-plane eigenenergies
which generally must be self-consistently calculated
\cite{sj,simserides:99,PL-theo,simserides-prb-2004,simserides-prb-2007-etc}.
Equation~(\ref{dos}) is valid for any kind of interplay
between spatial and magnetic confinement.
The $k_x$ dependence in Eq.~(\ref{dos}) increases
the numerical cost by a factor of $10^2-10^3$ in many cases;
hence it is sometimes overlooked,
although this is only justified for narrow single QWs or for $H \to 0$.
With the existing computational power,
such a compromise is not needed.
In the limit $H \to 0$, Eq.~(\ref{dos}) converges to
the staircase shape
with the famous step
$\frac {1}{2} \frac {m^* A}{\pi\hbar^2}$ for each spin.
The opposite asymptotic limit of Eq.~(\ref{dos}) is that of
a simple saddle point, where the DOS diverges logarithmically \cite{lyo:94}.
The DOS modification drastically affects the physical properties
\cite{sj,lyo:94,simserides:99,transport,plasmons-single,plasmons-double,PL-theo,PL-expe,simserides-prb-2004,
simserides-prb-2007-etc};
models which ignore it
can only be applied to narrow single QWs or for $H \to 0$.
For completeness, we note that in Eq.~(\ref{dos}) disorder is ignored.
With the progress of the epitaxial techniques
it is fairly small in well-prepared III-V structures.
Disorder will induce some broadening of the subbands.

The total population, $N$,
the internal energy, $U$,
the entropy \cite{shannon:48}, $S$,
and the free energy, $F$, are given by:

\begin{equation}\label{total_population}
N = \int_{-\infty}^{+\infty} \! d{\mathcal E}
\rho({\mathcal E}) f_0({\mathcal E}),
\end{equation}

\begin{equation}\label{internal_energy}
U = \int_{-\infty}^{+\infty} \! d{\mathcal E}
\rho({\mathcal E}) f_0({\mathcal E}) {\mathcal E},
\end{equation}

\begin{equation}\label{entropy}
S = -k_B \int_{-\infty}^{+\infty} \! d{\mathcal E}
\rho({\mathcal E}) f_0({\mathcal E}) ln[f_0({\mathcal E})],
\end{equation}

\begin{equation}\label{free_energy}
F = U - TS.
\end{equation}

\noindent $f_0(\mathcal E)$ is the Fermi-Dirac distribution function.
Since $\rho({\mathcal E}) \propto A$,
it follows that $N$, $U$, $S$ as well as $F$ are proportional to $A$.
However, the magnetization,

\begin{equation}\label{magnetization}
M = - \frac {1}{V} \Big( \frac {\partial F}{\partial B} \Big)_{N,T},
\end{equation}

\noindent where $V$ is the structure's volume, is independent of $A$.
To have the usual units in Tesla, we symbolize $B = \mu_0 H$,
$\mu_0$ is the magnetic permeability of free space.
To calculate $M$ we have to keep the temperature, $T$, as well as $N$ constant
e.g. assuming that all dopants are ionized. Here $T =$ 4.2 K.
Hopefully, below we list all the remaining symbols
used in the present article:
$N_i$ are the ``sheet'' (measured e.g. in cm$^{-2}$) subband concentrations, while
$N_s = N / A$ is  the ``sheet'' (measured e.g. in cm$^{-2}$) total concentration.
We use $E_i(k_x)$ for the subband energy dispersions
i.e. in the present article we ignore spin-splitting
which has been treated in detail elsewhere \cite{simserides-prb-2007-etc}.
Finally, for $H =$ 0, the ``the symmetric -asymmetric gap'',
$\Delta SAS = E_1(k_x=0) - E_0(k_x=0)$.

\begin{figure}[h!]
\includegraphics[height=5.0cm]{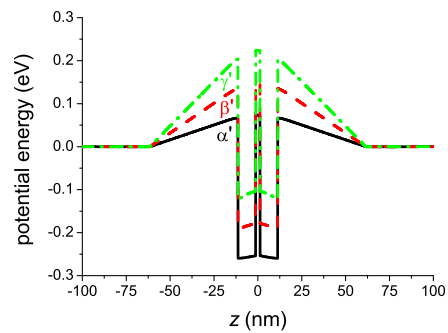}
\caption{(Color online) QW potential energy profiles, for $H =$ 0.
($\alpha '$) $ N_s = $ 1.85 $ \times $ 10$^{11}$ cm$^{-2}$,
$\Delta SAS = $ 4.24 meV.
($\beta '$) $ N_s = $ 3.70 $ \times $ 10$^{11}$ cm$^{-2}$,
$\Delta SAS = $ 4.01 meV.
($\gamma '$) $ N_s = $ 5.55 $ \times $ 10$^{11}$ cm$^{-2}$,
$\Delta SAS = $ 3.79 meV.
}
\label{fig:QW_profiles}
\end{figure}

\begin{figure}[h!]
\includegraphics[height=5.0cm]{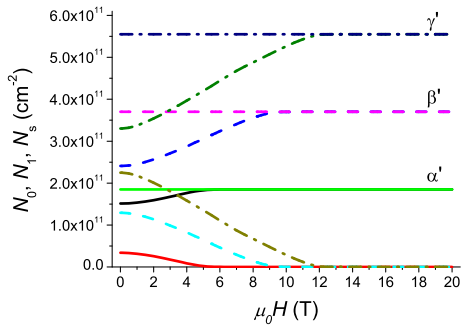}
\caption{(Color online) The sheet subband concentrations $N_0$, $N_1$
and the sheet total concentration $N_s$
as functions of the external in-plane magnetic field, $\mu_0 H$,
for cases $\alpha '$ (solid lines), $\beta '$ (dashed lines),
$\gamma '$ (dash-dotted lines).}
\label{fig:Ni-Ns}
\end{figure}

As a prototype system, we choose GaAs/(Al,Ga)As double QWs,
a bilayer system with well defined
``symmetric-asymmetric gap'' for $H =$ 0
and well-known material parameters.
Magnetization measurements under perpendicular magnetic field
of such a bilayer 2D system
with strong coupling between the two QWs
can be found elsewhere \cite{bominaar_et_al}.
To facilitate the reader we provide in Fig.~\ref{fig:QW_profiles}
the self-consistent potential energy profiles, for $ H = $ 0,
of the various double QWs employed in the present article.
Two (left and right) 50 nm spacers separate the $\delta-$doped layers
from the double QW.
The total double QW width is 22.7 nm including
the internal barrier of 2.5 nm.
For simplicity we take $A =$ 1 m$^2$.
Augmenting the $\delta$-doping,
we vary $N_i$, $N_s$ and $\Delta SAS$,
distinguishing three cases:
($\alpha '$)
$ N_0 = $ 1.51 $ \times $ 10$^{11}$ cm$^{-2}$
and
$ N_1 = $ 0.34 $ \times $ 10$^{11}$ cm$^{-2}$,
i.e.
$ N_s = $ 1.85 $ \times $ 10$^{11}$ cm$^{-2}$,
while $\Delta SAS = $ 4.24 meV.
($\beta '$)
$ N_0 = $ 2.41 $ \times $ 10$^{11}$ cm$^{-2}$
and
$ N_1 = $ 1.29 $ \times $ 10$^{11}$ cm$^{-2}$,
i.e.
$ N_s = $ 3.70 $ \times $ 10$^{11}$ cm$^{-2}$,
while $\Delta SAS = $ 4.01 meV.
($\gamma '$)
$ N_0 = $ 3.30 $ \times $ 10$^{11}$ cm$^{-2}$
and
$ N_1 = $ 2.25 $ \times $ 10$^{11}$ cm$^{-2}$,
i.e.
$ N_s = $ 5.55 $ \times $ 10$^{11}$ cm$^{-2}$,
while $\Delta SAS = $ 3.79 meV.
Fig.~\ref{fig:Ni-Ns} depicts $N_0$, $N_1$ and $N_s$
as functions of $\mu_0 H$ for all cases.
We notice that $N_s(\alpha '):N_s(\beta '):N_s(\gamma ') = 1:2:3$.
The depopulation of $E_1$ induced by the DOS modification
occurs approximately at 6 T, 10 T, and 12.5 T, respectively.

\begin{figure}[h!]
\includegraphics[height=5.0cm]{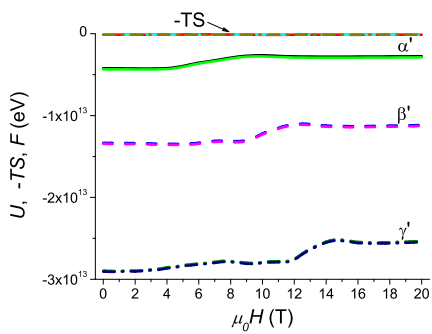}
\caption{(Color online) The internal energy, $U$,
the product $-TS$ and the free energy, $F$, as functions of
the external in-plane magnetic field, $\mu_0 H$,
for all cases ($\alpha '$, $\beta '$, $\gamma '$).
On this scale, $F \approx U$, $-TS$ is negligible.}
\label{fig:U-MTS-F}
\end{figure}

\begin{figure}[h!]
\includegraphics[height=5cm]{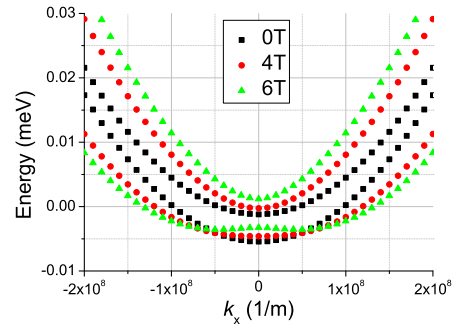}
\includegraphics[height=5cm]{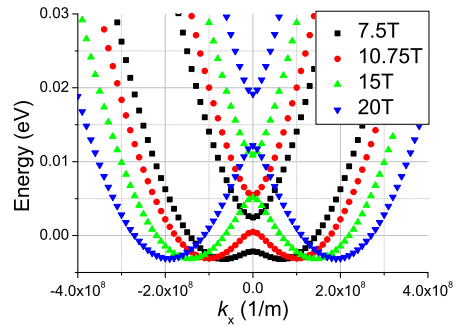}
\caption{(Color online)
Zooms of the energy dispersion, $E_i(k_x)$ ($i =$ 0, 1),
for characteristic values of $\mu_0 H$, for case ($\alpha '$).
Here the Fermi energy is identified with zero.}
\label{fig:alpha-dispersion}
\end{figure}

Fig.~\ref{fig:U-MTS-F} depicts $U$, $-TS$ and $F$,
as functions of $\mu_0 H$, for all cases.
At  $T = $ 4.2 K, $|F| \approx |U| \gg |-TS|$.
Since $N$ is kept constant in each case,
we expect that $|U|$ will decrease whenever $H$ induces
``flattening'' of the occupied subbands
i.e., expansion of the occupied parts to higher $|k_x|$,
since this leads to occupied energies with smaller $|{\mathcal E}|$.
The gradual increase of $|F|$ from ($\alpha '$) to  ($\beta '$) and ($\gamma '$)
mirrors the increase of the population.
To facilitate the reader we provide in Figs.~\ref{fig:alpha-dispersion}
zooms of the energy dispersion of case ($\alpha '$)
for characteristic values of $\mu_0 H$.

\begin{figure}[h!]
\includegraphics[height=5.0cm]{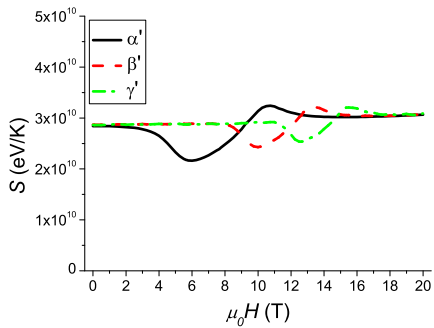}
\caption{(Color online) The entropy, $S$, as a function of $\mu_0 H$,
for each one of the cases ($\alpha '$, $\beta '$, $\gamma '$).}
\label{fig:S}
\end{figure}

Fig.~\ref{fig:S} depicts the entropy $S$ as a function of $\mu_0 H$
for all cases ($\alpha '$, $\beta '$, $\gamma '$).
According to Eq.\ref{total_population} and Eq.~\ref{entropy},
since for each case $N$ is constant,
$S$ is sensitive to the changes of $ln[f_0({\mathcal E})]$.
At $T =$ 4.2 K, these changes only occur in a short region
around the Fermi energy, $E_F$.
In other words, $S$ {\it reads} the modification of the energy dispersion
around $E_F \equiv$ 0.
For case ($\alpha '$), for $\mu_0 H =$ 0,
the bottom of $E_1(k_x)$ is relatively close to $E_F \equiv $ 0.
From 0 to 6 T,
$S$ is continuously falling
due to the continuous depopulation process of $E_1(k_x)$.
The minimum of $S$ occurs at 6 T because there
the gradual DOS modification completely depopulates the 1st excited subband,
cf. Fig.~\ref{fig:Ni-Ns} and Fig.~\ref{fig:alpha-dispersion}.
In crude language this means less anarchy, less entropy.
From 6 T to 10.75 T,
$S$ is continuously increasing because around $E_F \equiv$ 0,
$E_0(k_x)$ in the range $|k_x| \approx$ 0
moves upward and finally
the populated $E_0(k_x)$ is divided into two parts,
cf. Fig.~\ref{fig:alpha-dispersion}.
The maximum of $S$ occurs at 10.75 T because there
$E_0(k_x)$ splits completely into two occupied parts,
around $k_x =$ 0 the ground state subband ceases to be occupied.
Thus, the cohesion of the occupied $E_0(k_x)$ is lost around 10.75 T.
Hence, more anarchy, more entropy.
From 11 T to 20 T,
$S$ does not change much,
because the main effect of increasing $\mu_0 H$ is
to move the two $E_0(k_x)$ minima continuously apart,
cf. Fig.~\ref{fig:alpha-dispersion}.
The behavior of $S$, in cases ($\beta '$, $\gamma '$)
can be readily explained in analogous manner.
As one could maybe imagine,
increasing the system's magnitude, the minimum of entropy,
$S_{min}$, is increasing:
at     6 T with $S_{min} =$ 2.2 $\times 10^{10}$ eV/K,
at    10 T with $S_{min} =$ 2.4 $\times 10^{10}$ eV/K,
at  12.5 T with $S_{min} =$ 2.5 $\times 10^{10}$ eV/K.

\begin{figure}[h!]
\includegraphics[height=5.0cm]{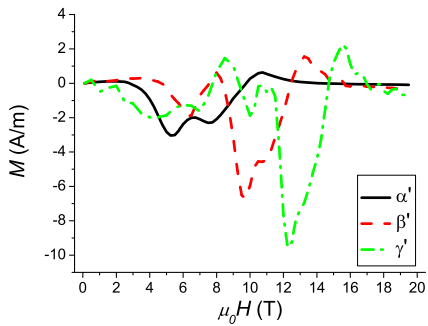}
\caption{(Color online) The magnetization $M)$ as a function of $\mu_0 H$,
for each one of the cases ($\alpha '$, $\beta '$, $\gamma '$).}
\label{fig:M}
\end{figure}

Fig.~\ref{fig:M} shows the magnetization $M$ as a function of $\mu_0 H$
for all cases ($\alpha '$, $\beta '$, $\gamma '$).
We observe that the DOS modification induces an oscillation of $M$; it is
between $\approx$ -3 A/m and $\approx$ 0.5 A/m for case ($\alpha '$),
between $\approx$ -7 A/m and $\approx$ 2   A/m for case ($\beta '$), and
between $\approx$ -9 A/m and $\approx$ 2   A/m for case ($\gamma '$).
The reader may observe that the magnetic susceptibility,
$\chi_m = \partial M / \partial H$,
swings between negative and positive values,
thus Fig.~\ref{fig:M} shows a totally orbital diamagnetic to paramagnetic fluctuation.
For example, for the case ($\gamma '$)
the fluctuation of $M$ of the order of 10 A/m,
is translated to approximately $\frac{1}{5}$
of the ideal de Haas-van Alphen effect
i.e. the magnetization step of two effective Bohr magnetons per electron
in the perpendicular configuration \cite{wilde_et_al_2006,peierls}.

In summary, the main message of this article is that
an in-plane magnetic field causes
a considerable effect
in the magnetization of quasi two-dimensional carriers
which has been ignored by the community up to now.
The magnetic susceptibility swings
from negative to positive values
i.e. we have proved that
an entirely orbital diamagnetic to paramagnetic fluctuation exists.
We conjecture that the in-plane component of a tilted magnetic field
will bring about similar effects,
hence care must be taken for the interpretation
of such magnetization measurements.
The entropy of quasi 2D carriers at low $T$ was also calculated and interpreted.

I thank Dr. I.~M.~A.~Bominaar-Silkens and Dr. U. Zeitler for useful discussions.



\begin{thebibliography}{18}

\bibitem{landau} L.~Landau, Z. Phys. {\bf 64}, 629 (1930).

\bibitem{IQHE} K.~von~Klitzing, G.~Dorda and M.~Pepper,
Phys.~Rev.~ Lett. {\bf 45}, 494 (1980).

\bibitem{FQHE} D.~C.~Tsui, H.~L.~St\"ormer and A.~C.~Gossard,
Phys.~Rev.~Lett. {\bf 48}, 1559 (1982).

\bibitem{laughlin} R.~B.~Laughlin, Phys. Rev. Lett. {\bf 50}, 1395 (1983).

\bibitem{dehaas-vanalphen} Early experimental efforts include e.g.
F.~F.~Fang and P.~J.~Stiles Phys.~Rev.~B {\bf 28}, 6992 (1983),
J.~P.~Eisenstein, H.~L.~St\"ormer, V.~Narayanamurti, A.~Y.~Cho,
A.~C.~Gossard, and C.~W.~Tu, Phys.~Rev.~Lett. {\bf 55}, 875 (1985).

\bibitem{shubnikov-dehaas} A.~B.~Fowler, F.~F.~Fang, W.~E.~Howard and P.~J.~Stiles, Phys.~Rev.~Lett. {\bf 16}, 901 (1966).

\bibitem{wilde_et_al_2006} M.~A.~Wilde, M.~P.~Schwarz, Ch.~Heyn, D.~Heitmann,
D.~Grundler, D.~Reuter and A.~D.~Wieck, Phys.~Rev.~B {\bf 73}, 125325 (2006).

\bibitem{peierls}R.~Peierls, Z. Phys. {\bf 81}, 186 (1933).

\bibitem{earlier}
S.~A.~J.~Wiegers, M.~Specht, L.~P.~L\'evy, M.~Y.~Simmons, D.~A.~Ritchie,
A.~Cavanna, B.~Etienne, G.~Martinez, and P.~Wyder, Phys.~Rev.~Lett. {\bf 79}, 3238 (1997);
J.~G.~E.~Harris, R.~Knobel, K.~D.~Maranowski, A.~C.~Gossard, N.~Samarth, and D.~D.~Awschalom,
Phys.~Rev.~Lett. {\bf 86}, 4644 (2001);
M.~P.~Schwarz, M.~A.~Wilde, S.~Groth, D.~Grundler, C.~Heyn, and D.~Heitmann,
Phys.~Rev.~B {\bf 65}, 245315 (2002);
M.~Zhu, A.~Usher, A.~J.~Matthews, A.~Potts, M.~Elliott, W.~G.~Herrenden-Harker,
D.~A.~Ritchie, and M.~Y.~Simmons, Phys. Rev. B {\bf 67}, 155329 (2003);
M.~A.~Wilde, M.~Rhode, C.~Heyn, D.~Heitmann, D.~Grundler, U.~Zeitler, F.~Sch\"affler,
and R.~J.~Haug, Phys.~Rev.~B {\bf 72}, 165429 (2005).

\bibitem{sj} L.~Smr\u{c}ka and T.~Jungwirth,
J. Phys.: Condens. Matter {\bf 6}, 55 (1994).

\bibitem{lyo:94} S.~K.~Lyo, Phys.~Rev.~B {\bf 50}, 4965 (1994).

\bibitem{simserides:99} C.~D.~Simserides, J.~Phys.: Condens.~Matter {\bf 11}, 5131 (1999).

\bibitem{simserides-prb-2004} C.~Simserides, Phys.~Rev.~B {\bf 69}, 113302 (2004).

\bibitem{transport}
J.~A.~Simmons, S.~K.~Lyo, N.~E.~Harff and J.~F.~Klem, Phys.~Rev.~Lett. {\bf 73}, 2256 (1994);
A.~Kurobe, I.~M.~Castleton, E.~H.~Linfield, M.~P.~Grimshaw, K.~M.~Brown,
D.~A.~Ritchie, M.~Pepper and G.~A.~C.~Jones, Phys.~Rev.~ B {\bf 50}, 4889 (1994);
T.~S.~Lay, X.~Ying and M.~Shayegan, Phys. Rev. B {\bf 52}, R5511 (1995);
T.~Jungwirth, T.~S.~Lay, L.~Smr\u{c}ka and M.~Shayegan, Phys.~Rev.~B {\bf 56}, 1029 (1997);
O.~N.~Makarovskii, L.~Smr\u{c}ka,
P.~Vasek, T.~Jungwirth, M.~Cukr, L.~Jansen,
Phys.~Rev.~B {\bf 62}, 10908 (2000).

\bibitem{plasmons-single} S.-J.~Cheng and R.~R.~Gerhardts, Phys.~Rev.~B {\bf 65}, 085307 (2002).

\bibitem{plasmons-double} S.~V.~Tovstonog and V.~E.~Bisti, JETP Letters {\bf 78}, 722 (2003).

\bibitem{PL-theo} D.~Huang and S.~K.~Lyo, Phys.~Rev.~B {\bf 59}, 7600 (1999).

\bibitem{PL-expe} M.~Orlita, R.~Grill, P.~Hl\'{\i}dek, M.~Zv\'ara G.~H.~D\"ohler,
S.~Malzer, M.~Byszewski, Phys.~Rev.~B {\bf 72}, 165314 (2005).

\bibitem{simserides-prb-2007-etc} C.~Simserides, Phys.~Rev.~B {\bf 75}, 195344 (2007);
C.~Simserides and I.~Galanakis,
{\it Proceedings of the 17th International Conference on
the Electronic Properties of Two-Dimensional Systems,
Genova, 2007}, Physica E, in press.

\bibitem{shannon:48} C. E. Shannon, Bell Syst. Tech. J. {\bf 27}, 379 (1948).

\bibitem{bominaar_et_al} I.~M.~A.~Bominaar-Silkens, U.~Zeitler, P.~C.~M.~Christianen,
D.~Reuter, A.~D.~Wieck, J.~C.~Maan, Physica E {\bf 34}, 191 (2006).

\end{thebibliography}
\end{document}